# The Importance and Criticality of Spreadsheets in the City of London

Grenville J. Croll
EuSpRIG – European Spreadsheet Risks Interest Group
mail@grenvillecroll.com

*Spreadsheets have been with us in their present form for over a quarter of a century. We have become so used to them that we forget that we are using them at all. It may serve us well to stand back for a moment to review where, when and how we use spreadsheets in the financial markets and elsewhere in order to inform research that may guide their future development. In this article I bring together the experiences of a number of senior practitioners who have spent much of their careers working with large spreadsheets that have been and continue to be used to support major financial transactions and manage large institutions in the City of London. The author suggests that the City of London is presently exposed to significant reputational risk through the continued uncontrolled use of critical spreadsheets in the financial markets and elsewhere.*

## 1 INTRODUCTION

The purpose of this paper is to give an overview of the uses to which spreadsheets are put in large organisations and major businesses operating within the City of London (an ancient yet well defined socio-economic area now including Canary Wharf). The aim is to make as explicit as is possible within the bounds of commercial confidentiality the reliance that is placed upon spreadsheets within one of the world's largest financial markets. The sources of the material are professionals from a variety of large organisations who have kindly granted the author telephone interviews under assurances of anonymity in order to permit a full and frank disclosure. Interviewees consisted of approximately twenty auditors, accountants, actuaries, bankers, directors, insurers, lawyers, quantitative finance specialists, IT specialists, regulators and other individuals playing significant roles in the City of London. The duration of each telephone interview (in each case conducted by this author) was approximately half an hour. The interviews were not recorded, however, hand written notes were taken. The research methodology was inspired by MacMillan [MacMillan, 2000]. Interviewees were called a few days after having received the title, abstract and most of this section by way of introduction. Further material was obtained from an internet discussion forum popular within the financial markets [Wilmott, 2005]. The author has collated and edited the contributions and provides a summary and conclusion.

Interviewees were informed that it was intended that each individual should characterise the uses of spreadsheets in their sector in terms of the people involved and their roles, the number of spreadsheets involved, frequency of use, their size, importance, significance or criticality. Significance or criticality could relate to finance, business opportunity, business risk, commercial necessity, fiduciary or legal duty, individual careers, health & safety etc. It was indicated that information on the professionalism with which spreadsheets are put together, perhaps in terms of Grossman's eight principles of Spreadsheet Engineering [Grossman, 2002] would be useful. Interviewees were informed that it was hoped that the information gathered from the exercise would complement the formal survey approach presently being undertaken by academic colleagues.

**2 THE CITY OF LONDON**

The Corporation of London on their website [CofL, 2005] introduce the City of London thus:

> "The City of London is the world's leading international financial and business centre…This sector made a net contribution to the UK's current account of over £13bn, a significant amount of which was generated within the Square Mile. The City contributes 3% to the UK's GDP"

The City of London's contribution to UK GDP pales besides its international financial significance [CofL, 2005]:

- $504bn foreign exchange turnover each day in London;
- 45% of the global foreign equity market
- 70% of Eurobonds traded in London
- $2,000bn per annum traded on metals in London
- London is the world's leading market for international insurance. UK worldwide premium income reached £161bn in 2002
- $884bn a day traded on the London international futures exchange
- £2,619bn total assets under management in the UK in 2002
- £1,046bn in overseas earnings generated by the maritime industry in London
- 287 foreign banks in London
- 19% of international bank lending arranged in the UK (largest single market)
- 381 foreign companies listed on the London Stock Exchange (LSE)
- £1,200bn pension fund assets under management (third largest in the world)
- $275bn daily turnover in 'over the counter' derivatives (36% of global share)

The main sectors given are Accountancy, Banking, Business information sources, Exchanges, Finance, Fund Management, Government, Insurance, Management Consulting, Maritime, Public Private Partnerships, Regulation, Transnational organisations and Transport. According to a recent survey [CEBR, 2005] commissioned by the Corporation of London, the City of London supports approximately 317,000 "City-Type" financial jobs. It would be safe to assume that all of these jobs required some familiarity with spreadsheets, most likely Microsoft Excel. We estimate that up to 30% of these jobs would be working with Important, Key or Critical spreadsheets as defined below.

**3 DEFINING IMPORTANCE AND CRITICALITY**

We originally defined importance in terms of the explicit financial value represented in or by the spreadsheet. An important spreadsheet was one that had lots of large financial numbers in it, where material error in those numbers could cause the organisation significant financial loss of the kind that has been well documented by Panko [Panko, 2000] et al.

A regulator, however, was kind enough to outline the mechanism by which they intend in future to measure the importance or criticality of spreadsheets submitted to them by regulated firms. Generalising their scale for wider use, it would seem sensible to categorise spreadsheets as follows:

| Type | Description and Potential Impact of errors |
|---|---|
| Critical | Material error could compromise a government, a regulator, a financial market, or other significant public entity and cause a breach of the law and/or individual or collective fiduciary duty. May place those responsible at significant risk of criminal and/or civil legal proceedings and/or disciplinary action. |
| Key | Material error could cause significant business impact in terms of incorrectly stated assets, liabilities, costs, revenues, profits or taxation etc. May place those responsible at risk of adverse publicity and at risk of civil proceedings for negligence or breach of duty and/or internal disciplinary action, |
| Important | Material error could cause significant impact on the individual in terms of job performance and career progression without directly, greatly, immediately or irreversibly affecting business or the organization. |
| Store & Retrieve | Spreadsheets used as databases, with few issues other than data correctness and information security and where the impact of error is low. |
| Expired | Spreadsheets over three years old no longer required in the active management of the business, but may be required to be archived by statute or good practice. Present impact of error is low. |
| Personal | Other spreadsheets used by the individual in the day to day performance of their duties, where the impact of material error is low. |

It was this regulator's intention that spreadsheets be annotated as above and with a similar confidentiality classification. There was a requirement for Critical spreadsheets that at least two people each side of the regulatory divide were aware of their function & purpose.

The above scale enables us to discuss more precisely the role of spreadsheets in the City of London.

**4 THE ROLE OF SPREADSHEETS IN THE CITY OF LONDON**

Three differing contributors to an internet forum [Wilmott, 2005] recently addressed the author's question regarding the present role of spreadsheets in the financial markets:

> *"Put simply and succinctly, despite the higher operational risk, Excel is everywhere - it is the primary front-line tool of analysis in the financial business. Most traders price deals in spreadsheets and enter them in large-scale deal capture systems afterwards"*

> *"Excel is utterly pervasive. Nothing large (good or bad) happens without it passing at some time though Excel"*
>
> *".. if a customer wants to do a bespoke trade that cannot be handled in our designated booking system then we have to book it in a spreadsheet and add a few bips to cover operational risk capital charge (and additional maintenance). How many bips to add is subjective"*

In a later telephone conversation, a regulator summarised thus:

> *"Spreadsheets are integral to the function and operation of the global financial system"*

When asked to comment upon the previous comment, a senior investment research professional working in a first tier investment bank commented:

> *"Agreed. In addition I would say that the majority of people who use the financial system are not appreciative of the role that spreadsheets play"*

Using the above as an introduction to the central role that spreadsheets play, it would seem pertinent to document in more detail their use in the main market sectors of the City of London. The contributions received from interviewees fell into the following categories: Professional Services, Law, Government, Financial Trading, Regulation, Investment Research, Investment Banking, Electricity & Gas, Rail, Private Finance, Fund Management, Insurance & Re-Insurance and Maritime. Although interviewees covered a broad spectrum, the following sections summarize a few major sectors, outlining the major issues.

**4.1 Professional services & Large corporate**

The four large professional services firms based in the City of London support the statutory reporting and financial management of most of the UK's larger companies. A larger number of smaller firms support the remaining large corporates and the next, smaller, tier of commerce. The firms have a substantial overseas clientele.

There is extensive use of spreadsheet based financial and strategic modelling within all professional services firms largely for and on behalf of their large corporate and institutional clients. Controls are in place to ensure that spreadsheets developed for clients are of good quality [Read, N. & Batson, J.,1999] and strategies are in place to manage errors [PWC, 2004]. An earlier paper [Croll, 2002] documents the typical steps involved in determining the correctness of models within the professional services environment. The total number of Key and Critical spreadsheets formally and independently checked for correctness by the professional service firms in the City of London does not presently exceed 1,000 on an annual basis. Many spreadsheets are Key or Critical:

> *"It is not unusual for millions of pounds of time investment to be reflected in a spreadsheet upon which critical reliance is placed"*

The independently checked Key & Critical spreadsheets support the types of transaction that are typically reported in the business sections of the quality daily newspapers and other business press and thus form a relatively small proportion of the total business transacted in the City of London.

Note that whereas the provisions of the Sarbanes Oxley act affecting US companies, are already in force, Basel II requirements affecting G10 banks and the 8[th] European Directive affecting listed European Corporates only start to come into effect in 2006.

Spreadsheets are not only methods of controlling operational risk (a key pillar of Basel II) but also are themselves a source of operational risk. Under Basel II, effective operational risk controls mean reduced regulatory capital, ineffective controls mean increased capital. Thus the effectiveness, accuracy and riskiness of spreadsheets is a matter the regulator should (and indeed does) take into account when setting regulatory capital.

Key and Critical spreadsheets developed internally by their large corporate clients do not appear to be subject to the same scrutiny as spreadsheets developed internally by the professional services firms themselves. Whilst regulators and insurers have indicated that large spreadsheets are in use in the management of large financial institutions and corporations, there is no evidence to suggest that any of them are subject to any form of quality control. A regulator has indicated that subject to approval of the criticality scale outlined earlier by the regulator's main board, Critical spreadsheets in *that regulators* industry will *in future* have to go through the full software development lifecycle.

A concern was expressed that during the consummation of a financial transaction, there was often an undue reliance placed upon the spreadsheet commissioned by the banks or the project sponsor which the professional services firm was building or auditing. There was no "independence of mind" if a single spreadsheet was being used, even if all the formulae had been checked as correct. This was reminiscent of some earlier work on the issues of interpretation [Banks, D. & Monday, A., 2002] and overconfidence [Panko, 2003] in the use of spreadsheets. Conversely, it was noted that sometimes spreadsheets did not form a major part of a decision, particularly where considerations such as political strategy or health and safety were also important.

It became clear that in certain industrial sectors, particularly Oil and Gas, the demographic profile of some of the key participants was such that certain knowledge, residing within complex spreadsheets was in danger of being lost as these people retired. A similar comment was aired within the fund management industry, where there is an active and competitive job market and people rapidly move on.

There is evidence that the use of strategic and financial modelling seen within the larger corporate is trickling down into smaller firms. Even in smaller transactions, spreadsheets were seen to be critical as they often implement or crystallise financial or operational benchmarks which post deal could precipitate default in the event of them not being set correctly.

**4.2 Financial Trading & Fund Management**

There is extensive use of spreadsheets within financial trading of all types. Spreadsheet models used in financial trading are often large and complex. Some models include monte-carlo simulation and optimization routines contained either explicitly within the spreadsheet or implicitly as part of a macro or function. Interviewee comments such as the following were typical:

> *"The whole industry is run on a spreadsheet"*

> *"Critical – couldn't run the bank without them"*

> *"Spreadsheets are the user interface into the deal"*

Spreadsheets are used to value financial instruments of all types and guide decisions on what to trade and when. They are the prime data source when details of a transaction are uploaded into an institution's main settlement, risk management or other financial systems. With certain types of specialist or complex transaction, the spreadsheet will remain the prime record.

> *"This has legal consequences, e.g. for compulsory recordkeeping under Companies legislation or FSA regulation, use of these records as evidence for proving disputed transactions, and the possibility of being required by a court to produce (and authenticate) such records after many years (in some cases, even substantially more than 6 years after their creation)".* [Reed, 2005]

Statutory financial information, including capital adequacy, is communicated to regulators and insurers using large and complex spreadsheets.

Spreadsheets are particularly heavily used in the more innovative and hence more recent parts of the market, particularly credit derivatives, an area of the market with a significant daily turnover. The global market in Credit Derivatives was expected to have reached $4.8 trillion by 2004 [BBA, 2005], with the City of London remaining the world's major centre with over half the global market share. Unsurprisingly, there is a "massive dependence" on the use of spreadsheets in interest rate and equity derivatives.

There is an active software vendor sector providing the financial markets with highly specialist pre-built or custom built spreadsheet based financial functions. This market was first established in the mid eighties, with one vendor now claiming 30,000 clients [MBRM, (2005)].

There is clear evidence that market participants are controlled at a strategic level by senior personnel using spreadsheet models. Errors were experienced and corrected within these strategic spreadsheet models, but no systematic error discovery processes took place.

With the exception of quantitative finance professionals working as software engineers, there is almost no spreadsheet software quality assurance or appreciation of the software development life cycle as it might relate to spreadsheets. Spreadsheets built using well engineered code libraries were inevitably tinkered with later by traders, sales people, analysts and other users in an uncontrolled fashion.

There is endemic denial of the extent of the use of spreadsheets within the financial markets at the highest level even though spreadsheets are very clearly, indeed obviously, being used to support large parts of corporate activity. The ease with which malfeasance and malpractice can occur in conjunction with the use of spreadsheets, and the risks relating thereto, was almost palpable.

> *"The known weakness in spreadsheet security has already led to some public domain incidents, notably the AIB affair"*

A managing director of a regulator (who was not an interviewee) is certainly aware of the risks of end-user computing:

> *"Are integrated systems a holy grail? Not necessarily. The PC triggered a revolution in financial services products. New complex products are very much built on PC computing capacity and could not have happened without it. And yet, senior management in many financial services firms regard the use of PCs as a temporary solution that needs at some point to be integrated – and somehow larger systems are the only solution. What this thinking demonstrates is not really a concern for productivity but rather system management, control and audit. This "large systems" vision of the world poses risks because as long as systems are run off temporary PC solutions and are not part of larger systems, all of the controls that are in place for larger systems are forgotten about when applied to smaller systems. It is important that companies apply many of the same controls and management over their fragmented systems as they apply to mainframe systems instead of concentrating efforts on migrating local computing power into a single managed system."*

An interviewee in a first tier bank reported that there was increasing interaction with corporate lawyers regarding spreadsheets and an increasing trend towards complexity, but there was an increasing trend in the last few years "towards better engineering within large complex spreadsheets". Whether the latter comment is applicable in less tightly regulated sectors such as Hedge Funds remains to be seen.

Spreadsheets also perform a critical role in the fund management industry. An actuary working in a very large fund reported that 90% of their working day was spent within the spreadsheet environment, and that this was typical. The daily profit and loss and balance sheet for a fund, performance measurement, a "great deal" of analysis, asset liability matching, portfolio optimisation (eg of a £20Bn fund) and extensive data manipulation all occur within the spreadsheet environment. Spreadsheets were felt to be indispensable for anything innovative.

There is evidence that in the financial markets, spreadsheets are operating at, close to, or even beyond the present technological limits of their size and/or complexity. There were several reports of the present 256 column limit unnecessarily limiting the number of instruments in a financial portfolio and constraining the level of detail in temporal models. There were reports of difficulties in spreadsheets over 50Mb, a size which is not at all uncommon. Note that spreadsheets >1Gb already exist.

There is very firm evidence from the author's interviews that people who create or modify spreadsheets in the financial markets are almost entirely self-taught.

**4.3 Private Finance Initiative and Public Private Partnerships**

The Private Finance Initiative in the UK is the means by which the government now arranges for schools, hospitals and prisons etc to be built, owned and operated. It accounts for approximately 12% of UK government capital spending. Since inception in the mid-nineties approximately 500-600 projects have reached financial close, with many more projects in progress. PFI capital spending is worth several tens of billions of pounds.

Spreadsheets play a central and critical role in this process as they bring together in one place the essential characteristics of the legal agreements for building, staffing, heating, lighting, securing, maintaining and otherwise owning and operating the project in exchange for the single annual charge – the Unitary Charge – that the government pays:

> *"There is no more basic question in business than how much one should charge for one's product or service. That question could not be answered in PFI projects without detailed spreadsheet based financial modelling"*

"The financial model" is a defined term in the legal agreements relating to PFI projects. Legal agreements specify what the financial model must do, who is responsible for it and how often it should be run or re-run over the 20-30 year project life. There is an issue regarding the ability to continue to run spreadsheets over a 30 year period. There is an aftermarket in the re-implementation of financial models to support the day to day management of the project post financial close.

The requirements specifications for spreadsheet based financial models in the UK's privatised rail industry were identified as being particularly good:

> *"The language used in the ITT's with regard to model specification sets a very high standard"*

Bids for rail privatization projects are non-compliant if they fail to meet the specified spreadsheet modelling standards. Within Rail privatisations, there has been a "significant elevation of the status of the model" within the last three years.

The spreadsheet modelling process in PFI (and other forms of limited recourse financing) is now relatively mature, as models and modelling methods have converged to support the generally consistent requirements of the various types of project. Spreadsheet models are extensively checked for correctness using cell-by–cell inspection and other methods [Ettema et al, 2001].

Approximately 1,000 people per year receive specialist training in the City of London in financial modelling skills relevant to PFI and investment banking – 0.3% of "City Type" Employees and 1% of "City Type" employees thought to be working with Important, Key or Critical spreadsheets.

**4.4 Law and Maritime**

The primary stock in trade of the Law is of course words rather than numbers, and this is reflected in the relatively low-key use of spreadsheet technology within the profession. Spreadsheets are in use where one might expect, in calculating personal injury claims or preparing accounts for example, but they do not play anything other than a peripheral role. A lawyer suggested that spreadsheets were often used to get the right answer ie to calculate the right assumptions to support significant conclusions already reached, a practice which the author has seen from time to time. There were 16,368 solicitors practicing in the City of London in 2003 [Law Society, (2003)].

Use of spreadsheets within Maritime is more widespread as they are used for data storage, data manipulation, the calculation and conversion of charter rates and tonne-mile costs. There is widespread use in the preparation of accounts and financial forecasts. The recent emergence of a shipping derivatives market has seen the very occasional use of spreadsheets for calculating Value-at-Risk using monte-carlo simulation. Maritime accounts for approximately 10,000 jobs in the City of London, across ship broking, ownership and insurance.

## 5 SUMMARY AND CONCLUSION

The intensity of spreadsheet use varies across the City of London as one would expect.

Within certain large sectors, spreadsheets play a role of such critical importance that without them, companies and markets would not be able to operate as they do at present.

In only a few of these market sectors are the risks relating to the use of spreadsheets relatively well controlled - in clearing banks, professional services, private finance and investment banking for example.

By contrast, in other sectors where spreadsheets play a Critical role, their use is relatively uncontrolled – the financial markets, fund management, investment research and financial reporting. In these sectors the real risks of material error are not known, and in some cases the risks are vehemently denied. In the worst instances, and supported by interviewee evidence, the existence or use of spreadsheets is concealed by senior management.

In sectors where spreadsheets play merely a Key role – in the internal management of large corporations for instance - spreadsheet use is almost completely uncontrolled. This directly and causally exposes large corporations to significant losses which regularly occur and are publicly reported [EuSpRIG, 2005].

By way of example, during the course of conducting the present 23 interviews, two interviewees each disclosed without prompting (as this was not a research objective) a current and recent instance of where material spreadsheet error directly involving or concerning them had led to adverse events involving "many tens of millions of pounds".

The situation is reminiscent of the time prior to the unexpected collapse of Long Term Capital Management (LTCM), where there was faith in the infallibility of Nobel Prize winning people and methodologies.

> *"At the beginning of 1998, the fund had equity of $5 billion and had borrowed over $125 billion — a leverage factor of roughly thirty to one. LTCM's partners believed, on the basis of their complex computer models, that the long and short positions were highly correlated and so the net risk was small."[LTCM, 2005]*

Spreadsheets have been shown to be fallible, yet they underpin the operation of the financial system. If the uncontrolled use of spreadsheets continues to occur in highly leveraged markets and companies, it is only a matter of time before another "Black Swan" event occurs [Taleb, 2001], causing catastrophic loss. It is completely within the realms of possibility that a single, large, complex but erroneous spreadsheet could directly cause the accidental loss of a corporation or institution, significantly damaging the City of London's reputation.

As it is impossible to imagine the City of London operating without the widespread use of large spreadsheets, it is important to encourage firms to identify, prioritise, test, correct and control Key and Critical spreadsheets. Doing so will enable firms to improve profitability by avoiding the inevitable occurrence of large spreadsheet related mistakes, of which only the smallest and least embarrassing are being reported.

Other actions that can be taken include: educating the market regarding the financial, regulatory and other risks of uncontrolled spreadsheet use; improving the regulatory framework regarding their use; promoting the emerging discipline of spreadsheet engineering; training management and staff to build & test more robust spreadsheet models; deploying software tools for detecting and correcting errors; deploying software tools to assist in the management of spreadsheets; encouraging the development of next generation spreadsheet technology and investigating other tools that have the advantages of spreadsheets but fewer or different disadvantages.

## 6 ACKNOWLEDGEMENTS

The author gratefully acknowledges the generous cooperation of the 23 interviewees without whom this paper would not have been possible. The author thanks a number of interviewees and the anonymous referee for their constructive comments.